\documentclass[10pt,conference]{IEEEtran}
\usepackage{amssymb}
\usepackage{amstext}
\usepackage{amsmath}
\usepackage{epsfig}
\usepackage{epic,eepic}

\newtheorem{theorem}{Theorem}[section]

\newcommand{\qed}{\nobreak \ifvmode \relax \else
     \ifdim\lastskip<1.5em \hskip-\lastskip
      \hskip1.5em plus0em minus0.5em \fi \nobreak
      \vrule height0.75em width0.5em depth0.25em\fi}

\begin{document}
\title{Interference Alignment on the Deterministic Channel and Application to Fully Connected AWGN Interference Networks}
\author{\authorblockN{Viveck Cadambe, Syed A. Jafar}
\authorblockA{Electrical Engineering and Computer Science\\
University of California Irvine, \\
Irvine, California, 92697, USA\\
Email: vcadambe@uci.edu,syed@uci.edu\\ \vspace{-1cm}}
\and
\authorblockN{Shlomo Shamai (Shitz)}
\authorblockA{Department of Electrical Engineering \\
Technion-Israel Institute of Technology\\
Technion City, Haifa 32000, Israel\\
Email: sshlomo@ee.technion.ac.il\\ \vspace{-1cm}
}}

\maketitle
\IEEEpeerreviewmaketitle

\begin{abstract}
An interference alignment example is constructed for the deterministic channel model of the $K$ user interference channel. The deterministic channel example is then translated into the Gaussian setting, creating the first known example of a fully connected Gaussian $K$ user interference network with single antenna nodes, real, non-zero and contant channel coefficients, and no propagation delays where the degrees of freedom outerbound is achieved. An analogy is drawn between the propagation delay based interference alignment examples and the deterministic channel model which also allows similar constructions for the $2$ user $X$ channel as well.
\end{abstract}
\section{Introduction}
Understanding the capacity of wireless networks is the ``Holy Grail'' of network information theory. Since exact capacity characterizations are unlikely to be found for most multiuser communication scenarios, there is an increased interest in approximate and/or asymptotic capacity characterizations as a means to understanding the performance limits of wireless networks. Promising approaches in this direction include degrees of freedom characterizations \cite{Jafar_dof,Cadambe_Jafar_int, Jafar_Shamai, Cadambe_Jafar_X} and deterministic channel models \cite{Avestimehr_Diggavi_Tse,Avestimehr_Diggavi_Tse_Flow,Bresler_Parekh_Tse}. Degrees of freedom characterizations seek the asymptotic scaling of network capacity with signal-to-noise-ratio (SNR). Deterministic channel models have lead to capacity characterizations within a constant number of bits for several interesting cases. While a precise connection between deterministic channel models and degrees of freedom characterizations has not been made in general, it is clear that the two approaches have a lot in common. Both the degrees of freedom perspective as well as the deterministic channel perspective focus on the high SNR regime and in both approaches, the noise is de-emphasized in order to gain a better understanding of the broadcast and interference aspects of wireless networks. In this paper, we explore further the relationship between these two perspectives.

One of the main results to come out of the degrees of freedom perspective is the concept of interference alignment. Interference alignment refers to the idea that signals can be designed to cast overlapping shadows at the receivers where they constitute interference while they remain distinguishable at the receivers where they are desired. This idea has lead to some surprising results for wireless networks. For example, it has been shown that a $K$ user interference network has $K/2$ degrees of freedom. In other words, as the total transmit power of the network is increased (or equivalently, as the AWGN power at each receiver is decreased), every user in an interference network will be able to simultaneously achieve half of the capacity (bits/sec/Hz) that he could achieve in the absence of the interference from other users. Similarly, for $M\times N$ node $X$ networks, i.e. networks of $M$ transmitters and $N$ receivers where an independent message needs to be communicated between each transmitter-receiver pair the number of degrees of freedom equals $\frac{MN}{M+N-1}$. Interference alignment is the key to this result as well. 

The degrees of freedom of wireless interference networks have been characterized under a variety of communication scenarios. However, several important questions remains open. One such unsolved question is to prove or disprove the Host-Madsen-Nosratinia conjecture \cite{Nosratinia_Madsen} that states that fully connected wireless interference networks with single antenna nodes and constant channel coefficients have only $1$ degree of freedom. The term ``fully-connected'' refers to the condition that \emph{all} channel coefficients are non-zero. In other words, all receivers see interference from all transmitters. For a fully connected $K$ user interference network the total degrees of freedom cannot be more than $K/2$. The achievability of $K/2$ degrees of freedom has been established for fully connected wireless interference networks under each of the following scenarios.
\begin{enumerate}
\item If the channels coefficients are chosen from a continuous distribution but allowed to vary over time or frequency slots, then the $K$ user fully connected interference network has $K/2$ degrees of freedom with probability $1$. The key is to treat multiple transmitted scalar symbols as a supersymbol, or a signal vector. The variations of the channel coefficients create a distinct linear transformation of the signal vectors between each transmitter receiver pair. Thus, the same set of transmitted signal vectors, after they pass through these distinct channels, are able to align at one receiver where they constitute interference and be distinct at another receiver where they are desired.
\item If the channel coefficients are chosen from a continuous distribution and held fixed and each node is equipped with $M>1$ antennas, then $MK/2$ degrees of freedom are achievable for $K=3$ users with probability $1$. In this case, the channel matrices provide the distinct spatial rotations that allow the signal vectors to align at one receiver and not align at another receiver.
\item It is shown in \cite{Cadambe_Jafar_int} that one can construct an example of a $K$ user fully connected interference network that achieves $K/2$ degrees of freedom with constant channel coefficients but the channel coefficient values are complex. Basically, the complex channel creates two dimensions so that interference can be aligned in one dimension (e.g. purely imaginary) while the desired signal is received in the other dimension (e.g. purely real).
\item It is shown in \cite{Cadambe_Jafar_int,AsilomarDelay} that one can construct an example of a $K$ user fully connected interference network that achieves $K/2$ degrees of freedom with constant channel coefficients by properly assigning propagation delays to the channels. However, because of the introduction of propagation delays this is not the classical interference channel model. The key is to have an even delay for desired transmitter receiver pairs and odd delays for undesired transmitter receiver pairs. If all transmitters send over even time slots, the choice of propagation delays ensures that at each receiver all interference is received over odd time slots and the desired signal is heard interference-free over even time slots.
\end{enumerate}
Interestingly, no example is known of a fully connected $K$ user interference network with constant and \emph{real} channel coefficients, with no delays and only a single antenna at each node, that can achieve $K/2$ degrees of freedom. This is the case even if we are allowed to pick the channel coefficient values. Intuitively, the difficulty is that we need a signal vector space where each channel provides a different transformation of the signal vectors. The signal vector spaces resulting from multiple channel uses (supersymbols) do not trivially solve the problem in this case because the channel stays constant across these supersymbols. Effectively each channel corresponds to a linear transformation that is a scaled identity matrix. Because of the multiplication with identity matrices, the signal vectors are not rotated, and their relative orientation is the same at all receivers. Thus, one cannot have the vectors align at one receiver and take distinct directions at another receiver. 

From the point of view of the deterministic channel approach, the possibility of interference alignment is quite intriguing as well. Interference alignment has been shown to be possible through lattice codes on a one-sided interference channel \cite{Bresler_Parekh_Tse}. On the one-sided interference channel all channel coefficients between transmitter $i$ and receiver $j$ are equal to zero unless, $j=1$ or $i=j$. Thus, the one-sided interference channel is not fully connected. To the best of our knowledge, no example is known so far where interference alignment is accomplished on a fully connected interference network even with the deterministic channel model. 

In this paper we explore the relationship between the deterministic channel model and the degrees of freedom perspective through the lens of interference alignment schemes. We accomplish the following objectives:
\begin{enumerate}
\item Provide an example of a deterministic model of a fully connected interference network where interference alignment is achieved so that each user is able to achieve half of the capacity that he would achieve in the absence of interference.
\item Translate the interference alignment example from the deterministic model into a fully connected Gaussian interference network with constant and \emph{real} channel coefficients, no delays and no multiple antenna nodes, that will achieve arbitrarily close to $K/2$ degrees of freedom.
\end{enumerate}

While the restriction to real channel coefficient values is somewhat artificial, the problem sheds light on the limits of the interference alignment concept. As it turns out this investigation also leads to a novel and rather surprising form of interference alignment, motivated by the deterministic channel but directly applicable to the fully connected $K$ user \emph{real} interference channel.

\section{Interference Alignment on the Deterministic Interference Channel}
\begin{figure}[h]
\centerline{\input{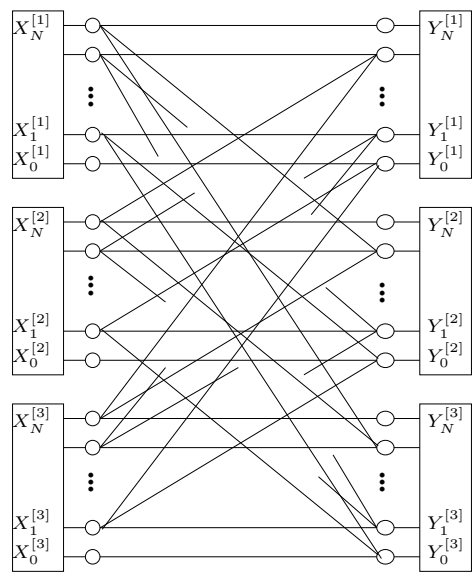}}
\caption{$3$ User Deterministic Interference Channel Example}\label{fig:detint}
\end{figure}

Figure \ref{fig:detint} illustrates a $3$ user fully connected interference channel model where interference alignment is accomplished in such a manner that each user achieves half of the capacity that he would achieve in the absence of interference. In this figure, transmitter $k$ ($k\in\{1,2,3\}$) sends $N+1$ bits that are numbered $X^{[k]}_0, X^{[k]}_1, X^{[k]}_N$ with the understanding that $X^{[k]}_N$ is the most significant bit. Receiver $k$ ($k\in\{1,2,3\}$) observes $N+1$ bits $Y^{[k]}_0, Y^{[k]}_1, Y^{[k]}_N$. On the deterministic channel, the channel only shifts the bits by an amount that depends on the SNR of that link \cite{Avestimehr_Diggavi_Tse, Avestimehr_Diggavi_Tse_Flow}. The bits that are shifted above the noise floor are the only bits that are received while the transmitted bits that end up below the noise floor at the receiver are lost. Note that at each receiver, the interference takes the form of an XOR (i.e. a modulo $2$ addition) of all the bits received at that level. In the example shown in Figure \ref{fig:detint} the outputs for the $k^{th}$ receiver ($k\in\{1,2,3\}$) are as follows:
\begin{eqnarray}
Y^{[k]}_N&=&X^{[k]}_N\nonumber\\
Y^{[k]}_i&=&X^{[k]}_i + \sum_{j\in\{1,2,3\}, j\neq k}X^{[j]}_{i+1}, ~~~i\in\{0, 1,2,\cdots,N-1\}\nonumber
\end{eqnarray}
In other words, the bits of the desired user are received without a shift, but the bits of the interfering users are received with a one bit shift. The interference alignment scheme works as follows. Let each transmitter set the value of all \emph{odd} bit positions as $0$. In other words:
\begin{eqnarray}
X^{[k]}_1=X^{[k]}_3=X^{[k]}_5=\cdots=X^{[k]}_N=0,
\end{eqnarray}
where we assume that $N$ is an odd number. Thus, no information is conveyed through these bits. The remaining bits (i.e. the bits in the even positions) are used such that each bit carries one bit of information. No error control coding is necessary for this deterministic example. Now, because of the shifts imposed by the channels in Figure \ref{fig:detint} it is easy to verify that:
\begin{eqnarray}
Y^{[k]}_i = X^{[k]}_i + 0 + 0 = X^{[k]}_i, \forall i\in\{0,2,4,\cdots,N-1\}
\end{eqnarray}
for all $k\in\{1,2,3\}$. In other words, each receiver has an interference-free channel to its desired transmitter over all the bits in even positions. Thus, each user is able to use half of their bits to convey information, effectively achieving a rate $(N+1)/2$ bits/s. Note that in the absence of interference, each user would achieve a rate $N+1$ bits/s. Thus, in this example, because of the interference alignment, each user achieves half of the capacity that they would achieve in the absence of interference. It is easy to see that this construction works for any number of users, i.e. $K>3$ as well. All we need is that the channel shifts the desired bits by an even amount and the interfering bits by an odd amount at each receiver.

Note the clear analogy between the deterministic channel construction and the delay example mentioned earlier. A $K$ user interference channel can achieve $K/2$ degrees of freedom if all desired links have an even propagation delay and all interfering links have odd propagation delay, where delay is measured in multiples of a basic symbol duration. Similar to the deterministic case, all transmitters stay silent over odd time slots and transmit only over even time slots and the result is that each receiver is able to hear its desired transmission free from interference over all the even time slots as the interference aligns itself over the odd time slots. The reason for this similarity is clear when we recognize that the shifting of the bits in the deterministic channel is very similar to the propagation delays on the real channel, where the shift happens in time. 

Next we return to the question of degrees of freedom of the fully connected $K$ user interference channel with constant and real channel coefficients, no delays and only single antennas at all nodes. We wish to translate the example presented above for the deterministic channel model into an example of interference alignment in the real case. We start with the channel model and the main result.

\section{The Gaussian Interference Channel}
We consider the $K$ user interference network described by the input-output equations:
\begin{eqnarray*}
Y^{[k]}(t)=\sum_{j=1}^KH^{[kj]}X^{[j]}(t)+Z^{[k]}(t)
\end{eqnarray*}
where at discrete time index $t$, $Y^{[k]}(t)$ and $Z^{[k]}(t)$ are the channel output and AWGN (respectively) at the $i^{th}$ receiver, $X^{[j]}(t)$ is the channel input symbol at the $j^{th}$ transmitter, and $H^{[kj]}$ is the scalar channel coefficient from the $j^{th}$ transmitter to the $k^{th}$ receiver. All symbols are real and the channel coefficients are fixed. The time index $t$ is suppressed henceforth for compact notation.  Note that we are only interested in constant channels as the channel coefficients are not a function of time.

For such a constant interference channel with real and non-zero coefficients, we wish to find out if interference alignment can be accomplished in a manner that $K/2$ degrees of freedom may be achieved. We explore this issue by constructing an interference channel with non-zero channel coefficients that can achieve within a factor $(1-\epsilon)$ of upperbound of $K/2$ degrees of freedom. In the process, we demonstrate a new kind of interference alignment scheme, that is inspired by the deterministic channel model, but applicable to the AWGN interference network that we consider.

We assume a transmit power contraint $P$ at each transmitter so that:
\begin{eqnarray*}
\mbox{E}\left[(X^{[j]})^2\right]\leq P.
\end{eqnarray*}
The AWGN is normalized to have zero mean and unit variance. $C(P)$ is the sum capacity of the $K$ user interference channel, and the degrees of freedom are defined as:
\begin{eqnarray}
d=\lim_{P\rightarrow\infty}\frac{C(P)}{\frac{1}{2}\log(P)}
\end{eqnarray}
The half in the denominator is because we are dealing with real signals only.

For this channel model we prove the following result.
\begin{theorem}
Given any $\epsilon>0$, there exists a fully-connected $K$-user Gaussian interference channel with constant and real coefficients that achieves $K/2(1-\epsilon)$ degrees of freedom.
\end{theorem}
\section{The Interference Alignment Scheme}
We will use representations of real signals in base $Q$ to construct the interference alignment scheme. The chosen value of $Q$ will be elaborated subsequently, but for now the reader may think of $Q$ as a finite but large and fixed integer multiple of $K$. We will represent real signals in base-$Q$ notation using $Q-$ary digits, which we denote as ``qits''. To avoid confusion we will mark the $Q$-ary representation as $[\cdot]_Q$.

Consider, for example the signal sent from transmitter $k$. We write:
\begin{eqnarray*}
X^{[k]}=\left[\cdots X_3^{[k]}X_2^{[k]}X_1^{[k]}X_0^{[k]}.X_{-1}^{[k]}X_{-2}^{[k]}X_{-3}^{[k]}\cdots\right]_Q
\end{eqnarray*}
where $X_i^{[k]}$  are, in general, integers with values between $0$ and $Q-1$, or equivalently the qits in the Q-ary expansion of the real number $X^{[k]}$. Equivalently, one may write:
\begin{eqnarray*}
X^{[k]}=\sum_{i=-\infty}^{\infty} X_i^{[k]}Q^i
\end{eqnarray*}

As a first step in the construction we pick the channel coefficients as follows:
\begin{eqnarray}
H^{[kj]}=\left\{\begin{array}{cc}1,&j=k\\Q^{-1},&j\neq k\end{array}\right.\label{eq:hkj}
\end{eqnarray}
Thus, all channel coefficients between transmitters and receivers that wish to communicate are equal to $1$ and all channel coefficients between interfering pairs of transmitters and receivers are equal to $Q^{-1} = [0.1]_Q$.

The transmitted symbol of user $j$ is constructed as follows:
\begin{eqnarray}
X^{[j]}&=&\left[\cdots X_3^{[j]}X_2^{[j]}X_1^{[j]}X_0^{[j]}.X_{-1}^{[j]}X_{-2}^{[j]}X_{-3}^{[j]}\cdots\right]_Q\\
&=&\left[X^{[j]}_{2N-2}0X^{[j]}_{2N-4}0X^{[j]}_{2N-6}0\cdots X^{[j]}_2 0 X^{[j]}_0.0\right]_Q
\end{eqnarray}
In other words $X^{[j]}_q = 0$ if $q\notin\{0,2,4,\cdots,2N-2\}$. The same construction is used for all transmitters. Thus, the only information in the transmitted symbol $X^{[j]}$ lies in the $N$ qits that are non-zero and occupy the even places above the decimal. 

Even though a base-Q representation allows all integer values between $0$ and $Q-1$ for the qits,  we place the constraint that all transmitted qits lie between $0$ and $\frac{Q}{K}-1$. This is done to avoid carry over from one qit to another when the interfering signals add at the receivers. Mathematically, 
\begin{eqnarray}
X^{[k]}_i&\in&\mathcal{X}\triangleq\{0,1,2,\cdots,\frac{Q}{K}-1\},\nonumber\\
 &&\forall k\in\{1,2,\cdots, K\}, \forall i\in \{0,2,4,\cdots,2N-2\}.\nonumber
\end{eqnarray}
In particular, our coding scheme will induce an i.i.d. uniform distribution over $\mathcal{X}$ on these $X^{[k]}_i, \forall k\in\{1,2,\cdots, K\}, \forall i\in \{0,2,4,\cdots,2N-2\}$.

Note that this construction establishes a functional relationship between $P$ and $N$ as:

\begin{eqnarray}
P &=& \mbox{E}\left[(X^{[j]})^2\right]\nonumber\\
&=& \mbox{E}\left[\left(\sum_{i=0}^{N-1}X^{[j]}_{2i}Q^{2i}\right)^2\right]
\end{eqnarray}
While one can explicitly calculate the exact dependence between $P$ and $N$, for the degree of freedom calculation it suffices to note that:
\begin{eqnarray}
\log_Q(P(N))= 4N+o(N)\label{eq:powerscale}
\end{eqnarray}
where $f(x)=o(g(x))$ denotes that $\lim_{x\rightarrow\infty}\frac{f(x)}{g(x)}=0$.
We focus on the scaling with $N$ rather than $Q$ because for the degrees of freedom innerbound in this paper, we fix $Q$ and let $N$ approach infinity. This is done so that the channel coefficients, which depends on $Q$ alone, are held constant and the transmit power $P$, which depends on $N$ as well, goes to infinity. Note that this is the typical setting for a degree of freedom characterization, i.e. the channel coefficients must be held fixed and only the power goes to infinity. 

Both the rates and the transmit power are expressed in terms of $N$ so that the following innerbound can be calculated. 
\begin{eqnarray}
d\geq\lim_{N\rightarrow\infty} \frac{\sum_{k=1}^KR^{[k]}(P(N))}{\frac{1}{2}\log_Q(P(N))}\label{eq:dofin}
\end{eqnarray}
where $R^{[k]}$ is the rate achievable by user $K$ with transmit power $P(N)$ under our achievability scheme. For the remainder of this section we will focus on the achievable sum rate.

The key idea for achieving $K/2$ degrees of freedom is interference alignment. What enables interference alignment in our setting is that multiplication by $Q^{-1}$ shifts the decimal point in the Q-ary representation $(X^{[j]})_Q$ by one place to the right. Thus, an interfering signal reaches a receiver shifted by one qit. For example, consider the signal from transmitter 2 as received at receiver 1:
\begin{eqnarray}
H^{[12]}X^{[2]}=\left[X^{[2]}_{2N-2}0X^{[2]}_{2N-4}0X^{[2]}_{2N-6}0\cdots X^{[2]}_2 0. X^{[2]}_00\right]_Q
\end{eqnarray}
Except for transmitter 1, every other transmitter's signal goes through this shift as it reaches receiver 1. Because of the shift, the information carrying qits of the desired signal $X^{[1]}$ are aligned with the zero padding bits of all interfering signals $H^{[1j]}X^{[j]}$ where $j\neq 1$. The precise details of the construction and the degrees of freedom calculation are presented next.

Because of the symmetry of the construction, we focus without loss of generality on Receiver $1$. The received signal at Receiver $1$ is expressed as:
\begin{eqnarray}
Y^{[1]}&=&\sum_{i=0}^{N-1}X^{[1]}_{2i}Q^{2i} + \sum_{i=0}^{N-1} \left(\sum_{k=2}^{K}X^{[k]}_{2i}\right)Q^{2i-1}+ Z^{[1]}\nonumber
\end{eqnarray}

Let us define an artificial signal $\overline{Y}^{[1]}$ which is a noise-free version of the actual received signal.
\begin{eqnarray}
\overline{Y}^{[1]}&=&\sum_{i=0}^{N-1}X^{[1]}_{2i}Q^{2i} + \sum_{i=0}^{N-1} \left(\sum_{k=2}^{K}X^{[k]}_{2i}\right)Q^{2i-1}\nonumber
\end{eqnarray}
Note that because the maximum value of a transmitted qit is only $\frac{Q}{K}-1$, the addition of the interference qits in $\overline{Y}^{[1]}$ does not produce a carry over. In other words, the addition of real interference signals is exactly equivalent to the modulo $Q$ addition of the corresponding qits. However, in the actual received signal $Y^{[1]}$ because the noise qits can take all values upto $Q-1$, the presence of noise can produce carry overs. As we argue next this problem disappears as we increase $N$ and consider the more significant qits.

Note that noise power is fixed at unity while $N$ grows to infinity. As $N$ increases, the disparity between the noise power and the power carried by the more significant qits increases.  Therefore, regardless of whether the noise takes positive or negative values, its impact diminishes as we consider the more significant bits. It follows then that:
\begin{eqnarray}
\lim_{N,i\rightarrow\infty, i\leq 2N-2} \mbox{Prob}\left(Y^{[1]}_i\neq\overline{Y}^{[1]}_i\right)=0.\label{eq:noiseoff}
\end{eqnarray}

Let us denote by $R^{[1]}_i$, the rate achieved by user $i$, by encoding in time over the qits $X^{[1]}_i(t)$. Note that for odd values of $i$, $C^{[1]}_i=0$ because these qits are forced to take the value $0$ by our construction. For even values of $i$, note that there is no interference from other users because of the interference alignment. Moreover it follows from (\ref{eq:noiseoff}) above, that the noise is also not an issue as $i$ increases. Thus,
\begin{eqnarray}
\lim_{N, i\rightarrow\infty, i\leq N-1}R^{[1]}_{2i} = \log_Q\left(\frac{Q}{K}\right)=1-\log_Q(K) ~~\mbox{qits/s}.
\end{eqnarray}
By symmetry the same arguments can be made for each user and we have:
\begin{eqnarray}
\sum_{k=1}^KR^{[k]}(P(N))=NK(1-\log_Q(K)) + o(N) ~\mbox{qits/sec}\label{eq:ratescale}
\end{eqnarray}

Substituting (\ref{eq:powerscale}) and (\ref{eq:ratescale}) into (\ref{eq:dofin}) we have
\begin{eqnarray}
d&\geq& \lim_{N\rightarrow\infty}\frac{NK(1-\log_Q(K))+o(N)}{\frac{1}{2}\left(4N+o(N)\right)} \\
&=& \frac{K}{2}(1-\log_Q(K))\\
&\geq&\frac{K}{2}(1-\epsilon)
\end{eqnarray}
for $Q>K^{1/\epsilon}$. Thus, for any given $\epsilon$ we are able to construct a $K$ user interference channel with constant channel coefficients that can achieve $\frac{K}{2}(1-\epsilon)$ degrees of freedom.

We conclude this section with an observation about the choice of the channel coefficients in (\ref{eq:hkj}). Note that for the interference alignment achieved in this section, it also suffices if we pick 
\begin{eqnarray}
H^{[kj]}=\left\{\begin{array}{cc}\alpha_{kj}Q^{n_e^{[kj]}},&j=k\\\alpha_{kj}Q^{n_o^{[kj]}},&j\neq k\end{array}\right.
\end{eqnarray}
where $n_e^{[kj]}$ and $n_o^{[kj]}$ are any even and odd integer values, respectively. In other words it suffices if the channel shifts the inputs by an even amount on the desired links and by an odd amount on the interference links (or vice versa). Note that it does not matter whether $n_e^{[kj]}$ and $n_o^{[kj]}$ are positive or negative integers. Thus, one can create examples where either the desired or the interference channels are stronger. In both cases examples can be constructed that achieve within $\epsilon$ of $K/2$ degrees of freedom. One can also accommodate coefficients $\alpha_{kj}$ that are not equal to $1$ and still choose $Q$ large enough that no carry overs are produced and the $K/2$ degrees of freedom can be approached within any desired $\epsilon$. 
\section{Conclusion}
We proposed an interference alignment example for the deterministic model of the $K$ user interference channel. The deterministic model leads to the first known example of a fully connected $K$ user Gaussian interference channel where all channel coefficients are real, non-zero constants and where $K/2-\epsilon$ degrees of freedom may be achieved for any $\epsilon>0$. The example is also interesting as it shows explicitly how the deterministic channel model translates to coding schemes for the real Gaussian interference channel at high SNR.

Another aspect of the example that is interesting is the analogy between the deterministic channel and the propagation delay example previously proposed in \cite{Cadambe_Jafar_int, AsilomarDelay}. The shifting of the qits in the deterministic channel model is analogous to the propagation delay which shifts the signal in time. In both cases, using an even/odd shift construction one is able to create interference channel scenarios that achieve the outerbound on the degrees of freedom. For the delay example, it is interesting to note that even if the delays are chosen randomly from a continuous distribution, one can ensure, with probability one, that the even/odd delay configuration needed for interference alignment is realized by choosing the basic symbol duration small enough  \cite{AsilomarDelay}. Thus, in the delay example, $K/2$ degrees of freedom are achieved with probability one, even with random delays. The analogy between delay and deterministic channel shifts hints at the interesting possibility that it may be possible to use a similar argument on the deterministic channel, and hence on the real Gaussian interference channel as well. 

Lastly, notice that delay based examples have been constructed for the 2 user $X$ channel as well \cite{Cadambe_Jafar_X}. These examples can also be easily converted into deterministic channel examples and by similar arguments into fully connected Gaussian X channel with real coefficients. Thus we also have the first known example of a fully connected X channel with single antenna nodes and real non-zero and constant channel coefficients where the upperbound on the degrees of freedom is achieved.
\bibliographystyle{ieeetr}
\bibliography{Thesis}

\begin{thebibliography}{1}

\bibitem{Jafar_dof}
S.~A. Jafar, ``Degrees of freedom for distributed mimo communications,'' in
  {\em IEEE Communication Theory Workshop}, June 2005.

\bibitem{Cadambe_Jafar_int}
V.~Cadambe and S.~Jafar, ``Interference alignment and the degrees of freedom of
  the k user interference channel,'' in {\em Preprint available through
  http://newport.eecs.uci.edu/~syed}, July 2007.

\bibitem{Jafar_Shamai}
S.~Jafar and S.~Shamai, ``Degrees of freedom region for the {MIMO} {X}
  channel,'' in {\em arXiv:cs.IT/0607099v3}, May 2007.

\bibitem{Cadambe_Jafar_X}
V.~Cadambe and S.~Jafar, ``Degrees of freedom of wireless {X} networks,'' in
  {\em Preprint available through http://newport.eecs.uci.edu/~syed}, 2007.

\bibitem{Avestimehr_Diggavi_Tse}
A.~S. Avestimehr, S.~Diggavi, and D.~Tse, ``A deterministic approach to
  wireless relay networks,''
\newblock Oct 2007, arXiv:cs.IT/0710.3777.

\bibitem{Avestimehr_Diggavi_Tse_Flow}
A.~S. Avestimehr, S.~Diggavi, and D.~Tse, ``Wireless network information
  flow,'' Sep 2007.
\newblock Allerton Conference.

\bibitem{Bresler_Parekh_Tse}
G.~Bresler, A.~Parekh, and D.~Tse, ``Approximate capacity of the many-to-one
  interference channel,'' Sep. 2007.
\newblock Allerton Conference.

\bibitem{Nosratinia_Madsen}
A.~Host-Madsen and A.~Nosratinia, ``The multiplexing gain of wireless
  networks,'' in {\em Proc. of ISIT}, 2005.

\bibitem{AsilomarDelay}
S.~Jafar and V.~Cadambe, ``{ Degrees of Freedom of Wireless Networks - What a
  difference delay makes},'' in {\em Asilomar Conference on Signals, Systems,
  and Computers, Pacific Grove, CA}, Nov 2007.

\end{thebibliography}
\end{document}